\documentclass[aps,prl,amsmath,twocolumn]{revtex4} %,twocolumn
\newcommand\R{\zeta}
\renewcommand{\section}[1]{\paragraph{\textbf{#1}}}
\renewcommand{\section}[1]{\noindent\paragraph{\bf\emph{#1}}}

\usepackage{xcolor}

	\usepackage{amsmath}%,amssymb} 
	\usepackage{amsfonts}
  	\usepackage{amssymb}
	\usepackage{makeidx}
	\usepackage{amsfonts}
	\usepackage[ansinew]{}
	\usepackage[usenames,dvipsnames]{pstricks}
	\usepackage{epsfig}
    \usepackage{graphicx}
	\usepackage{float}

\newcommand{\be}{\begin{equation}}
\newcommand{\ee}{\end{equation}}
\newcommand{\bea}{\begin{eqnarray}}
\newcommand{\eea}{\end{eqnarray}}
\newcommand{\ba}{\begin{align}}
\newcommand{\ea}{\end{align}}

\newcommand{\dd}{{\mathrm{d}}}

\makeindex

\begin{document}

\title{Sound speed induced production of primordial black holes}

\author{Antonio Enea Romano$^{1,2}$}

\affiliation{
${}^{1}$Theoretical Physics Department, CERN, CH-1211 Geneva 23, Switzerland\\
${}^{2}$ICRANet, Piazza della Repubblica 10, I--65122 Pescara, Italy \\
%${}^{3}$Instituto de Fisica, Universidad de Antioquia, A.A.1226, Medell\'{\i}n, Colombia \\
}
\begin{abstract}
    We study different mechanisms by which the speed of primordial curvature perturbations can produce an enhancement of the curvature spectrum, which could lead to the production of primordial black holes (PBH). One possibility is the growth of the sound speed in single scalar field models, which can induce super-horizon growth of curvature perturbations. The other is the momentum dependent effective sound (MESS) of curvature perturbations, which can arise in multi-fields models or modified gravity theories. 
    Future gravitational waves observatories such as LISA will allow to set constraints of the space and  time dependency of the sound speed, and on the different theoretical scenarios from which it can originate.
\end{abstract}

%\pacs{98.80.Es, 98.65.Dx, 98.80.-k}

\maketitle

\section{Introduction}
Primordial curvature perturbations are of fundamental importance in the standard cosmological model, since they provide the seeds from which it originated anything we can observe today, including for example large scale  structure (LSS) or  the anisotropies of the cosmic microwave background (CMB) radiation.

For slow-roll inflationary models comoving curvature perturbations $\R$ are known to be conserved on super horizon scales \cite{Lyth:2004gb}, a phenomenon which can be related to the vanishing of non adiabatic pressure perturbations $\delta P_{nad}$ on those scales.
Nevertheless, for generic single scalar field systems, the vanishing of non adiabatic pressure perturbations is not a sufficient neither necessary condition for the  conservation of $\R$, since  in globally adiabatic (GA) models \cite{Romano:2015vxz,Romano:2016gop}  $\delta P_{nad}=0$ on any scales, but  $\R$ may not freeze on super horizon scales. In a similar 
4
 way also a temporary phase of violation of slow roll conditions \cite{Vallejo-Pena:2019lfo,Garcia-Bellido:2017mdw,Romano:2008rr,Arroja:2011yu,Romano:2014kla,Cadavid:2015iya,GallegoCadavid:2016wcz} can induce an enhancement of the curvature spectrum, through the same mechanism of GA models, i.e. through a sudden change in the time evolution of the Universe expansion, causing a large growth of the conversion factor between entropy and curvature perturbations \cite{Vallejo-Pena:2019lfo}, while entropy perturbations remain small.

In this paper we will study two general mechanism which can induce the enhancement of the curvature spectrum, based on behavior of the sound speed.
In particular we will consider single field models with a growing sound speed, and multi-field or modified gravity theories which can induce a momentum dependence of the effective sound speed (MESS) \cite{Romano:2020oov,Romano:2018frb} of comoving curvature perturbations.
%In the GA models the curvature spectrum enhancement is due the super horizon growth of what in slow-roll models would be a decaying mode, due to a very fast decrease of the slow-roll parameter $\epsilon$.  Note that a similar behavior of $\R$ can happen in models which are not GA, such as for example in models with features of the inflaton potential \cite{}, in which cases the conversion factor between $\delta P_{nad}$ and $\R$ can have a large variation related to the sudden decrease of $\epsilon$.
Another mechanism is the growth of curvature perturbations due to the MESS, a quantity which allows to study the effects of entropy perturbations on curvature, and to study in a model independent way different physical scenarios such as multi-fields systems or modified gravity. In order to find a model independent estimation of the effect of the MESS, we derive an analytical formula to for the PBHs fraction produced by a local variation of the MESS.

%There exist another independent mechanism which can induce the super horizon growth of $\R$ in GA models, related to the growth of the sound speed $c_s$. 
%By imposing the condition $c_s \propto a^n$ we obtain the Lagrangian of one these models and show explicitly that the super horizon growth of $R$ is not due to the sudden decrease of $\epsilon$, but to the the growth of $c_s$ after the mode have crossed the horizon.
\section{Globally adiabatic models}
There exist two definitions of non adiabaticity in the literature of cosmological perturbations. One is based on the pressure perturbations in the uniform density gauge $\delta P_u=\delta P_{\delta\rho=0}$ \cite{Gordon:2000hv}
\be
\delta P_u=c_w(t)^2 \delta\rho+\delta P_{nad} \,,
\ee
where the the adiabatic sound speed is defined as $c_w=P'/\rho'$, and the subscript $u$ stands for uniform density gauge. Note that by construction the above definition is gauge invariant, and for this reason no gauge is specified in r.h.s of the last equality.
The other definition of entropy is based on the decomposition of pressure perturbations in the comoving slices gauge \cite{Kodama:1985bj} according to
\be
\delta P_c=c_s(t)^2 \delta \rho_c +\delta P_c^{nad} \,,
\ee
where $c_s(t)$ is the sound speed of comoving curvature perturbations $\R$, and the subscript $c$ stands for comoving slices  gauge.
Note that the above definition of $c_s$ is not unique as show in \cite{Romano:2018frb}, but for GA models this is not an issue since they satisfy the condition $\delta P_c^{nad}=\delta P_u^{nad}=0$ on any scale, which is a sufficient and necessary condition for  $c_s^2=c_w^2$, that is the defining property of  GA models \cite{Romano:2016_GA}. Note that the condition $c_s^2=c_w^2$ for perturbed perfect fluids implies that the equivalent scalar field system has a constant potential, i.e. $V(\phi)=V_0$, i.e. these models could  alternatively be defined as shift symmetric in the language of field theory. %This is the apprach adopted for example in constant roll inflationary models, which are special cases of GA models normally defined by the equation of motion of a shift symmetric scalar field.
\section{Background equations and adiabatic sound speed}
From  the  Friedmann's and continuity equations 
\bea
H^2=\frac{\rho}{3 M_p^2} \, &;& \dot{H}= -\frac{\rho +  P}{2 M_p^2} \, , \\
\dot{\rho}+3 H(\rho+P)&=&0 \,,
\eea
and the definition of the slow-roll parameter $\epsilon=-\dot{H}/H^2$,  we get 
\bea
\epsilon&=&\frac{3(\rho+P)}{2\rho} \label{epsilon} \,, \\ 
\rho'+\frac{2\epsilon \rho}{a}&=&\rho'+\frac{b}{a}  =0 \,, \label{cont} \eea
where a prime denotes derivatives with respect to the scale factor $a$ and $b=2\epsilon \rho$ is a useful quantity \cite{Romano:2016gop} to study the properties of cosmological perturbations.
The above system of three equations contains four functions $b,\epsilon,P,\rho$, implying that once one function has been fixed all the others can be obtained.

From the above equations we can get also the adiabatic sound speed in terms of $b(a)$
\bea
c^2_w =\frac{P'}{\rho'}&=&-1  -\frac{a b'(a)}{3 b(a)} \label{cb}\,.
\eea
All the above equations are completely general and can be applied to any homogeneous flat  Universe, but those  quantities in general are not enough to determine the evolution of comoving curvature  perturbations $\R$, which also requires to specify the sound speed $c^2_s$. In the special case of GA models we have  $c^2_s=c_w^2$, so that the evolution of $\R$ is completely determined by specifying one of those four functions.
In particular we will construct GA  models with a sound speed of the form $a^n$, since this could give rise to super-horizon growth of curvature perturbations.

\section{Super-horizon evolution of comoving curvature perturbations}
From the equation for the curvature perturbations on comoving slices,
\be
\frac{\partial}{\partial t}\left(\frac{a^3\epsilon}{c_s^2}
\frac{\partial}{\partial t}\R\right)-a\epsilon \Delta \R=0 \label{eomR2} \,,
\ee
after using the scale factor as a time variable, we get that on super horizon 
scales, i.e. on scales where the Laplacian term is negligible, beside the constant solution there is another one, corresponding to $\frac{a^3\epsilon}{c_s^2}
\frac{\partial}{\partial t}\R  \propto const $, given by \cite{Romano:2016_GA}
\bea
\R &\propto& \int^a\frac{da}{a} f(a)\,;
\quad 
f(a)\equiv{\frac{c_s^2(a)}{Ha^3 \epsilon(a)}} \,. \label{Rca}
\eea
When  $c_s^2$ and $\epsilon$ are
both slowly varying in time, the integral reaches quickly a constant value due to the $a^3$ suppressing effect, which is the reason why this solution is called decaying mode, causing the freezing of $\R$ on super-horizon scales.

When $\epsilon \propto a^n$, the  with $n \leq -3$ the normally "decaying" mode becomes a "growing" mode and $\R$ is not conserved on super horizon scales, which is the case for example of generalized ultra slow roll (GUSR) inflation and Lambert inflation \cite{Romano:2016_GA}. Note that for these models the sound speed is exactly constant for Lambert inflation or approximately constant in the case of GUSR, so the growth of $\R$ is entirely due to $\epsilon$. 
Here instead we will consider  the case when $c_s^2 \propto a^n$ and construct the corresponding Lagrangian.
\section{Scalar field models with growing sound speed}

The quantity $b(a)$ is convenient to find the Lagrangian of a GA model from the behavior of its background functions because of the  important relation \cite{Romano:2016gop}
\bea
X(a) &\propto& a^6  b(a)^2 \,.
\eea
where $X=g^{\mu\nu}\partial_\mu\phi\partial_\nu\phi$.
For a given $b(a)$ we can invert the above equation to get $a(X)$, and than from $P(a)$ we can finally get $K(X)=P[a(X)]$.

In general the relation $X(a)$ may not be inverted analytically, making it difficult to find a general analytical form of the Lagrangian corresponding to $c_s \propto a^n$. Even without an explicit form of the Lagrangian those models could be studied using the scale factor as a time variable, as long as the system of four equations (\ref{epsilon}-\ref{cont},\ref{cb}) can be solved analytically. 

\begin{figure}[H]
\centering
\includegraphics[width=.4\textwidth]{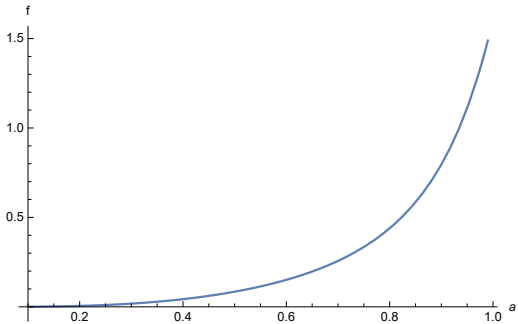}
\caption{The function $f$ is plotted as a function of the scale factor.  According to eq.(\ref{Rca})
\label{f} this kind of behavior implies the super-horizon growth of the what would be a decaying mode in a single field model with standard kinetic term.}
\end{figure}

\begin{figure}[H]
\centering
\includegraphics[width=.4\textwidth]{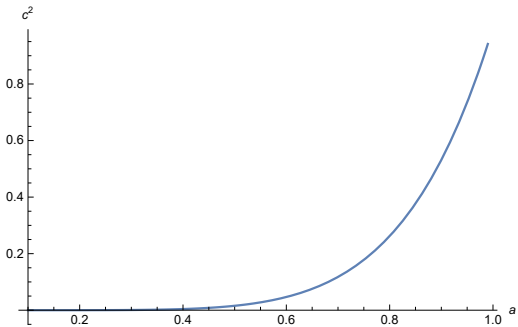}
\caption{The squared adiabatic sound speed is plotted as a  $c^2_w$ is plotted as a function of the scale factor. The adiabatic sound sound speed $c_w^2$ and the comoving curvature perturbations sound speeds $c_s^2$ are equal, and grow as $c_s^2=c_w^2 \propto a^6$. }
\label{c2}
\end{figure}

\begin{figure}[H]
\centering
\includegraphics[width=.4\textwidth]{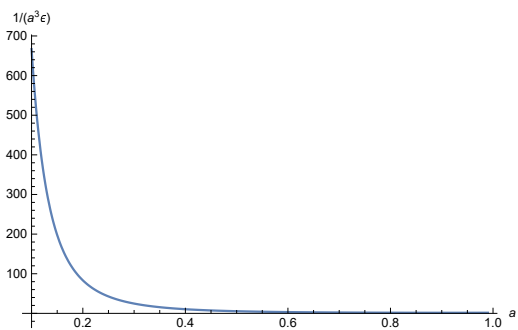}
\caption{The quantity $1/(a^3 \epsilon)$ is plotted as a function of the scale factor. Since this is a decreasing function of the scale factor, it is not the cause of the growth of the function $f(a)$, and according to eq.(\ref{Rca}), it is not the cause of the super-horizon growth of $\R$.}
\label{1Oae}
\end{figure}

In this paper we will consider a model for which the Lagrangian can be computed analytically, corresponding to $c_w^2 \propto a^6$, in which case from eq.(\ref{cb}) we get 
\bea
b \propto  \frac{e^{-\frac{a^6}{2}} }{a^3} \,&,&\, X \propto e^{-a^6} \,, \label{bx}
\eea
which substituted in eqs.(\ref{epsilon}-\ref{cont}), following the a procedure similar to the one described in \cite{Romano:2016_GA} gives
\bea
L&=&V_0-\text{erf}\left(\sqrt{\log \left(\frac{1}{\sqrt{X}}\right)}\right) \\
&=&V_0+K(X) \label{LX} \,,
\eea
where the constants in eq.(\ref{bx}), and in $c_w^2$ have been fixed appropriately to get the above form of the Lagrangian.
Note that, since $X$ is exponentially decreasing function of the scale factor, the argument of the square root in the kinetic term of the Langrangian is always positive for  appropriate initial conditions, corresponding to a positive sound speed. 
The slow roll parameter $\epsilon=-\dot{H}/H^2$ can be computed exactly as a function of the scale factor 
\bea
\epsilon&=&\frac{3}{6 e^{\frac{9 a^6}{\pi }} a^3 \left[\text{erf}\left(\frac{3 a^3}{\sqrt{\pi }}\right)-V_0\right]+2}  \,
\eea
from which we can see that the condition for inflation $\epsilon<1$ is easily satisfied.
In order to study the evolution of $\R$ on super horizon scales we plot the functions $f(a)$, $1/\epsilon a^3$ and $c_s^2(a)$ in figs.(\ref{f},\ref{c2},\ref{1Oae}). As shown in fig.(\ref{1Oae}) the growth of the function $f(a)$ is not due to $\epsilon(a)$, since $1/\epsilon a^3$ is a decreasing function, implying that the super horizon growth of $\R$ is entirely due to the sound speed. For this reason this model is fundamentally different from other  GA models exhibiting super-horizon growth of $R$ \cite{Romano:2016_GA}. In fact in those cases the growth was due the background evolution inducing a violation of slow-roll conditions, due to a fast decrease of $\epsilon$, while in these models the system is in slow-roll regime, but curvature perturbations growth because the sound speed is growing.
% \section{Globally adiabatic models}
% Globally adiabatic (GA) models are single field models defined by the property \cite{Romano:2016gop} of having vanishing non adiabatic pressure $\delta P_{na}$ perturbations on any scale
% \be
% \delta P_{na} &=&\delta P-c^2_w \delta\rho
% \ee
% where $c^2_w=\dot{P}/\dot{\rho}$ is the adiabatic sound speed.
% For $K(X)$ theories this implies that the sound speed coincides with the adiabatic sound speed $c^2_s=c^2_w$, which is also equivalent to the condition that the potential is constant.

.

\section{Transient growth of the sound speed}
A background model defined by the function $K(X)$ in eq.(\ref{LX}) gives rise to an exponential growth of the curvature perturbations which would cause a break down of the perturbative approach, and would be incompatible with observations, in particular CMB. A viable mode requires  that the system is described by that Lagrangian only for a limited time interval, during which the modes leaving the horizon will keep growing, according to eq.(\ref{Rca}), as $\R \propto a^3$.
For this reason a model compatible with observations could be described by this kind of Lagrangian
\be
L=a X+b e^{-\left( \frac{X-X_c}{\sigma} \right)^{2 q} } K(X)+V(\phi)
\ee
with $q>1$, and $a \ll b$, so that for $X \approx X_c$ the Langrangian is dominated by the $K(X)$ term, while for other values of $X$ the system would follow a slow roll evolution, for an appropriate choice of the potential $V(\phi)$. In this paper we will not study in details the properties of such a model, but we just outlined how an observationally viable could be constructed, and leave to a future work the study of the constraints of the parameters, and the PBHs production.
\section{Momentum dependent sound speed}
For any physical system satisfying gravitational field equations of the form $G_{\mu\nu}=T^{eff}_{\mu\nu}$, in absence of anisotropy, it is possible to derive an equation for comoving curvature perturbations  \cite{Romano:2018frb} 

\begin{align}
\ddot{\R}_k + \frac{\partial_t(Z_k^2)}{Z_k^2} \dot{\R}_k & + \frac{v_k^2}{a^2} k^2 \R_k = 0  \, , \label{Rckeq}  
\end{align}
where $ Z_k^2\equiv\epsilon a^3/v_k^2$, the dots denote derivatives with respect to cosmic time, and  the momentum dependent effective sound speed (MESS) $v_k(t)^2$ is defined according to
\begin{equation}
v_k^2(t) \equiv \frac{\alpha_k(t)}{\beta_k(t)} \, . \label{ck}
\end{equation}

Assuming that $v_k$ is not time dependent the above equation takes the form
\begin{align}
\R_k''+\frac{\partial_{\eta}(z^2)}{z^2}\R_k'+v_k^2k^2\R_k &=0 \, , \label{Rcpk} \\
u_k''+\left(v_k^2 k^2 -\frac{z''}{z} \right)u_k&=0 \, , \label{ukp}
\end{align}
where $u= z\R$ is the Sasaki-Mukhanov variable, and the primes denote derivatives with respect to conformal time. The MESS is encoding the effects of entropy on the curvature perturbations, and can be computed for any system with a well defined energy momentum tensor, including multi-fields \cite{Romano:2020oov}, and modified  gravity theories \cite{Vallejo-Pena:2019hgv}, after taking to the r.h.s. of the Einstein's equations the tensor due to the modification of gravity.
Due to its generality  the MESS approach is quite useful to study in a model independent way the effects of entropy on curvature perturbations.
An approximate solution of  eq.(\ref{ukp}), to leading order in slow-roll, can be written as
\be
u_k(\eta)=\frac {e^{-i v_k k \eta}}{\sqrt{2 v_k k }} \left(1-\frac{i}{ v_k k \eta}\right) \label{usol} \,.
\ee
Assuming that the above solution describes the  mode dominating the spectrum of curvature perturbations, which for multi-fields models, if possible, may require the choice of an appropriate field basis \cite{Romano:2020oov},
we can compute the spectrum of curvature perturbations as 
\be
P_{\R}=\frac{P^{0}_{\R}}{v_k^3} \label{Pvk} \,,
\ee
where $P_0$ is the spectrum without MESS, i.e. with $v_k-1$.

\section{PBHs produced by a peak in the curvature spectrum due to the MESS}
The present fraction $f_{PBH}$ of dark matter in the form  of PBHs of mass M can be approximated as \cite{Sasaki:2018dmp}
\begin{align}
    f= 2.7 \times 10^8 \left(\frac{\gamma}{0.2}  \right)^{1/2} \left(\frac{g_{*F}}{106.75}  \right)^{-1/4} 
     \left(\frac{M}{M_{\odot}}  \right)^{-1/2} \beta \, \nonumber ,
\end{align}
where $g_{*F}$ is the number of relativistic degrees of freedom at formation. The quantity $\beta$ is  the energy density fraction of PBHs at formation time
\begin{equation}
    \beta \equiv \frac{\overline{\rho}_{PBH}}{\overline{\rho}} \Bigr|_{F} \, ,
\end{equation}
which is related to the probability of  the density contrast $P(\delta)$ as \cite{Carr1975,Green:2004wb}
\begin{equation}
    \beta(M) = \gamma \int_{\delta_{t}}^{1} P(\delta) \dd \delta \, ,
\end{equation}
where   $\delta_t$ is the threshold for PBH formation. Assuming  a Gaussian distribution for the density perturbations, $\beta$ can be approximated as 
\begin{equation}
    \beta(M) \approx \frac{\gamma}{\sqrt{2\pi}\nu(M)} \exp\left[ -\frac{\nu(M)^2}{2}\right] \, ,
\end{equation}
where $\nu(M)\equiv \delta_t/\sigma(M)$, $\sigma(M)$ is the standard deviation  of the density contrast on scale $R$. The quantity $\sigma(M)$ is the link between PBHs formation and the curvature spectrum, to which it can be related using the Poisson's equation, giving
\begin{align}
    \sigma^2(M) &= \int \dd \ln k W^2(k R)\mathcal{P}_{\delta}(k) \nonumber \\ &= \int \dd \ln k W^2(k R) \left( \frac{16}{81}\right)(k R)^4\mathcal{P}_{\R}(k) \, , 
\end{align}
where  $W(k R)$ is the window function smoothing over the comoving scale $R(M)=(a^2 \mathcal{H})^{-1}\Bigr|_{F}=2 G M/a_{F}\gamma^{-1}$.
Note that the above  approximations to estimate the PBHs abundance can  receive  important corrections depending on the shape of power spectrum, and on the non- gaussianity \cite{Germani:2018jgr,Germani:2018jgr}.
As an example of the application of the  general formula obtained for the curvature spectrum in eq.(\ref{Pvk}), we can consider an  effective sound speed of the form
\be
v_k=v_0e^{\left(\frac{k-k_0}{ \sigma^2}\right)^{2/3}} \,,
\ee
which would produce a narrow peak in the primordial curvature perturbations spectrum similar to the one studied in \cite{Cai:2018dig}. Note that this is only a local approximation around $k_0$, and does not model accurately the power far from that scale, since it would be not relevant for the PBHs production.
Since in general the width of the local variation of the MESS can be negligible with respect to the scale $k_0$, i.e.  $\sigma \ll k_0$, which also implies $R \ll \sigma$,  the leading order contribution is given by  
\bea 
\beta&=&\frac{8 \sqrt{\pi }}{81 v_0^3}
\left(1+(R \, \sigma)^4 P_0 \right) 
\left(
\frac{k_0^3}{\sigma^3}+
\frac{3 k_0}{2 \sigma} 
\right) \,.
\eea

This prediction can be used to constrain in a model independent way primordial entropy perturbations produced in multi-fields systems or modified gravity theories, using    gravitational waves observations, and the corresponding constraints on  the abundance of the primordial black holes (PBHs).

The advantage of the MESS approach is that it allows to find a simple model independent representation of the effects of primordial entropy, which is expected to arise in any multi-field inflationary model \cite{Romano:2020oov}. Once the MESS of different models has been computed, it can be compared to the observational constraints obtained from using model independent parametrizations such as the one above.
With respect to a a purely phenomenological parametrization of the curvature spectrum, as the one adopted in \cite{Cai:2018dig} for example, the MESS allows to identify the fundamental physical origin the curvature spectrum shape, while still allowing a model independent analysis, as shown in \cite{Rodrguez:2020hot}.

Note that due the assumption of time independency of $v_k$ which was made to derive eq.(\ref{ukp}), according to eq.(\ref{Rca}), on super horizon scales there should still be a freezing of curvature perturbations, so the super-horizon evolution is not the cause of the enhancement of the curvature spectrum, which on the contrary is due to a modification of the Bunch-Davies vacuum, as can be seen in eq.(\ref{usol}).
% \begin{align}
%  \dot{\R} &= -\frac{v_s^2}{a^2 H \epsilon} \SPD\Psi_B - \frac{1}{3H\epsilon } \SPD\Pi \, . \label{RPhiBPi} 
% \end{align}

% \begin{align}
% \dot{\R} &= - \frac{c_s^2}{a^2H\epsilon} \SPD\Phi_B - \frac{\Gamma}{2H\epsilon} - \frac{1}{3H \epsilon} \SPD \Pi\, , \label{RPhiGamma} \\ 
% \ddot{\R}&+\frac{\partial_t z^2}{z^2}\dot{\R}-\frac{c_s^2}{a^2}\SPD\R + \frac{c_s^2}{\epsilon}\SPD \Pi  + \frac{1}{z^2}\partial_t \left[ \frac{a^3}{c_s^2 H} \left(\Gamma + \frac{2}{3}\SPD \Pi \right) \right] = 0 \, , \label{Rdotcgamma}
% \end{align}

% \be
% v_s^2=c_s^2\left(1-\frac{\Gamma}{\alpha}\right)^{-1}
% \ee
\section{Conclusions}
We have identified two different mechanisms by which the speed of primordial curvature perturbations can produce an enhancement of the curvature spectrum, which could lead to the production of primordial black holes.  One possibility is the growth of the sound speed in single scalar field models, which can induce super-horizon growth of curvature perturbations. As an example, the Lagrangian of a single scalar field model with sound speed growing as $a^6$ has been derived. 

We have then considered the effects of the momentum dependent effective sound of curvature perturbations, which can arise in multi-fields models or modified gravity theories, showing how it can provide a model independent explanation for the peaks of the curvature spectrum, which can be constrained by future gravitational waves observations.

In this paper we have considered separately the effects of the time and the momentum dependency of the MESS, but in the future it could be interesting to consider cases where they are both important, such as in certain multi-fields models \cite{Romano:2020oov}.

\section{Acknowledgments}
We thank Juan Garc\'ia-Bellido and Jose Mar\'ia Ezquiaga for useful comments and discussions. 

\bibliographystyle{h-physrev4}
\bibliography{mybib}
\end{document}